\documentstyle[bo99,epsfig]{article}

\title{NEW EXTREME SYNCHROTRON BL LAC OBJECTS }
\author{L. Costamante$^{1,2}$, G Ghisellini$^2$, P. Giommi$^3$, 
G. Tagliaferri$^2$, A. Celotti$^4$, M. Chiaberge$^4$, 
L. Chiappetti$^5$, G. Fossati$^6$, L. Maraschi$^2$, E. Pian$^7$,
F. Tavecchio$^2$, A. Treves$^8$, A. Wolter$^2$ } 
\affil{1) Universit\`a degli Studi di Milano, Milano, Italy;
2) Osservatorio Astronomico di Brera, Milano, Italy;  
3) BeppoSAX Science Data Center, ASI, Roma, Italy; 
4) S.I.S.S.A., Trieste, Italy;
5) IFCTR/CNR, Milano, Italy;
6) CASS/UCSD, La Jolla, California, USA;
7) TESRE, Bologna, Italy;
8) Universit\`a dell'Insubria, Como, Italy \\  }

\begin{document}

\maketitle

\begin{abstract}
We report on the {\footnotesize {\it Beppo}}SAX observations of four ``extreme" BL Lacs,
selected to have high synchrotron peak frequencies. 
All have been detected also in the PDS band.
For 1ES 0120+340, PKS 0548--322 and H 2356--309 the  
spectrum is well fitted  by a convex  broken power-law, thus locating the
synchrotron peak around 1--4 keV. 
1ES 1426+428 presents a  flat energy spectral index ($\alpha_x=0.92$) 
up to $\sim$100 keV, thus constraining the synchrotron peak to lie near or above
that value.
For their extreme properties, all sources could be strong TeV emitters.

\end{abstract}

\section{Introduction }
BL Lac objects are usually divided in two main classes, on the basis of
their overall Spectral Energy Distribution (SED): LBL or HBL (low or high
 energy peaked BL Lacs), according as the peak of the synchrotron emission 
(in a $\nu F_{\nu}$ representation) is in the
 IR--optical or EUV--soft-X band, respectiveley.
In the X-ray band this usually translates in a spectral index which is steep 
($\alpha_x>1$) for HBLs (corresponding to the tail of the synchrotron emission)
and  flat ($\alpha_x<1$) for LBLs (corresponding to the upcoming of the 
Inverse Compton emission).
In 1997, the {\it Beppo}SAX observations of Mkn 501 (Pian et al. 1997) 
and 1ES 2344+514 (Giommi et al. 1997) revealed that, at least in a 
flaring state, the peak of the  synchrotron emission can actually reach
very high energies, around 100 keV, 
with a consequently flat synchrotron X--ray spectral index. 
In order to find and study other sources with such ``extreme" properties,
we have selected several candidates from the Einstein Slew Survey and the
Rosat All Sky Survey Bright Sources Catalogue (RASSBSC).
%
\begin{figure}[t]
\centerline{\psfig{file=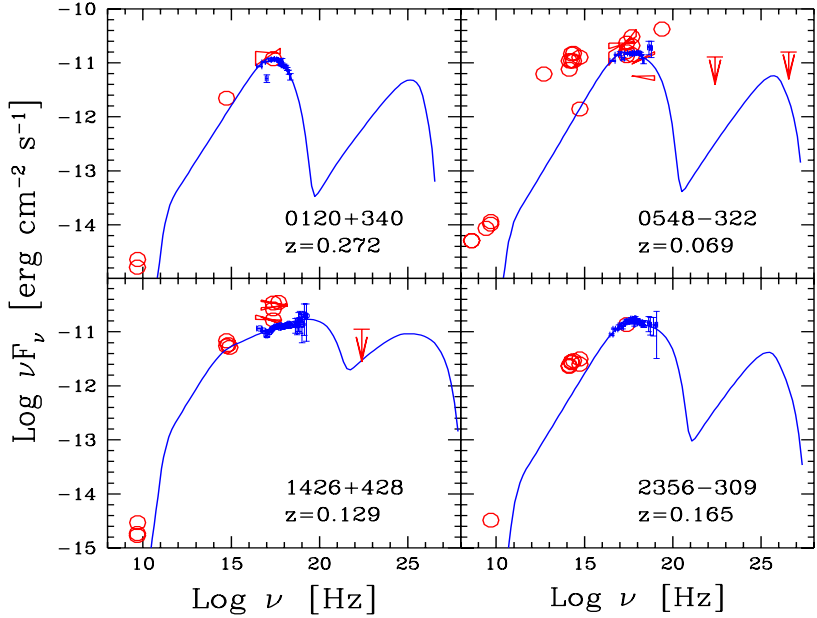, width=11cm, height=9.8cm}}
\vspace*{-15mm}
\caption[]{\footnotesize The SEDs of the 4 BL Lacs, made with {\it Beppo}SAX and literature 
data,  together with a pure
homogeneous SSC model (further details in Costamante et al., in prep.)}
\end{figure}
%
The selection criteria were based on properties  suggesting a high 
$\nu_{peak}$: \\
a) very high $F_x / F_{radio}$ ratio ($>3\times10^{-10}$  erg cm$^{-2}$
s$^{-1}$ / Jy,  at [0.1--2.4] keV and 5 GHz respectively); 
b) flat X-ray spectrum (when available), connecting smoothly with 
the flux at lower frequencies; 
c) appropriate values of $\alpha_{ro}$, $\alpha_{ox}$ and $\alpha_{rx}$
(Padovani \& Giommi 1995). A high X--ray flux ($>10^{-11}$ erg cm$^{-2}$ 
s$^{-1}$) in the 2--10 keV band  was also requested, to achieve a good
detection  in the PDS instrument. \\
We used the {\it Beppo}SAX satellite, whose wide X--ray energy range
(0.1--200 keV) is ideal to constrain the synchrotron peak.
Four objects have been observed, between June 1998 and April 1999:
1ES 0120+340, PKS 0548--322, 1ES 1426+428 and H 2356--309.
In Fig. 1 and Table 1 
only the best fit results are reported 
(for a complete discussion, see Costamante et al.,
in preparation). 
LECS, MECS and PDS data have been reduced and analysed according to
the SDC Cookbook instructions, using the latest calibration matrices available.
 Standard extraction radii of $4^{\prime}$ and $8^{\prime}$ for MECS and LECS
were used, except for the PKS 0548--322 observation of 20/2/99: in this case
a $6^{\prime}$ radius for the LECS has been used, due to the presence of a
contaminating source in the field of view (identified  as 
the star \verb+GSC_07061_01558+ in the Guide Star Catalog, probably flaring). \\
The PDS instrument doesn't have imaging capabilities, and its f.o.v.
(radius $\sim45^{\prime}$) is larger than LECS and MECS ($\sim28^{\prime}$ 
for the MECS). Therefore there is the possibility for PDS spectra to be
contaminated by hard serendipitous sources in the f.o.v, not visible
in the MECS images. Analyzing PDS data, we have taken this into account,
also checking in the NED database for potentially contaminating sources.
%
%
\section{1ES 1426+428}
At $41^{\prime}$ from this source, thus in the PDS f.o.v., there is the quasar 
GB 1428+422 
(Fabian et al 1998). To account for its contribution, we have added a 
component to the
PDS model, based on the GB 1428+422 data from the {\it Beppo}SAX observation  
of 4/2/99 
($\alpha= 0.42$, F$_{1keV}= 0.30\; \mu$Jy;  Celotti \& Iwasawa, priv. comm.).  
We have also checked in the NED and WGACAT databases for other potentially
contaminating objects: we added the contributions of the two
most important objects (WGA J1426.1+4247 and CRSS 1429.7+4240, Fig. 2 right 
panel), according to the fluxes and spectral indices 
extrapolated from the ROSAT band (when data were not available in literature, 
we used galactic N$_{\rm H}$ and a HR--$\alpha$ conversion by Giommi, 
priv. comm.).
Summing all  components, the different off--axis response of the instrument
has been taken into account. The PDS/MECS normalization has 
been fixed at 0.9. \\
With this model (Fig. 2 right panel), adding the PDS data to the 
LECS+MECS fit yields a $\chi^2_r= 1.06$, with the PDS points still
slightly above the model (Fig.2 left upper panel). 
A better $\chi^2_r$ (0.95) is obtained with GB1428+422 flux as a free 
parameter: in this case the resulting flux is F$_{1keV}= 1.44\pm0.54$,
a factor more than 4 higher during this observation
than the week before. 
{\it Anyway, in both cases, the spectrum of 1ES 1426+428 remains 
flatter than unity up to 100 keV.} 
%
\begin{figure*}
\begin{tabular}{rl}
\hspace*{-2.5cm} 
{\psfig{file=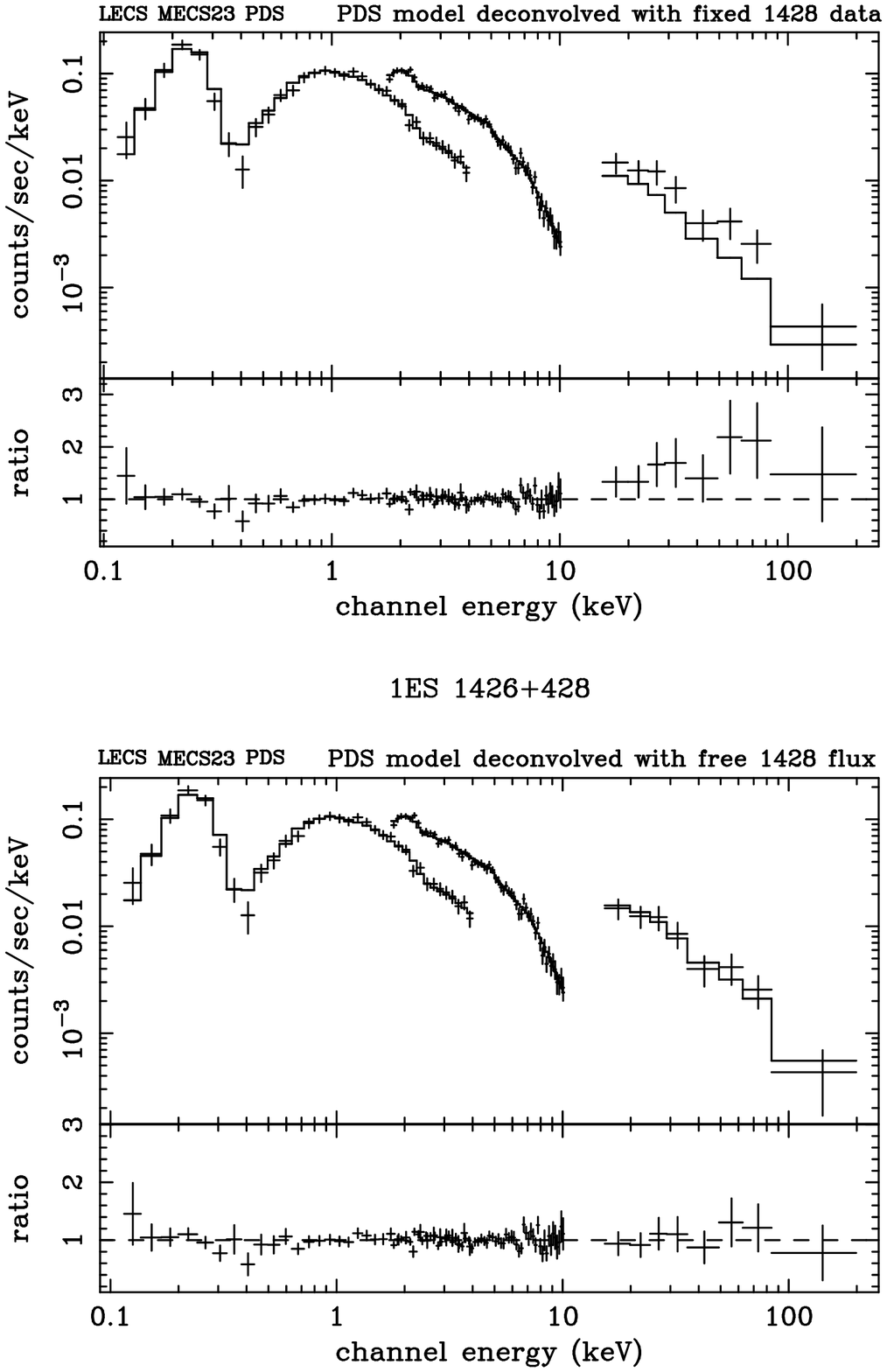, width=7cm, height=10cm }}&{\psfig{
file=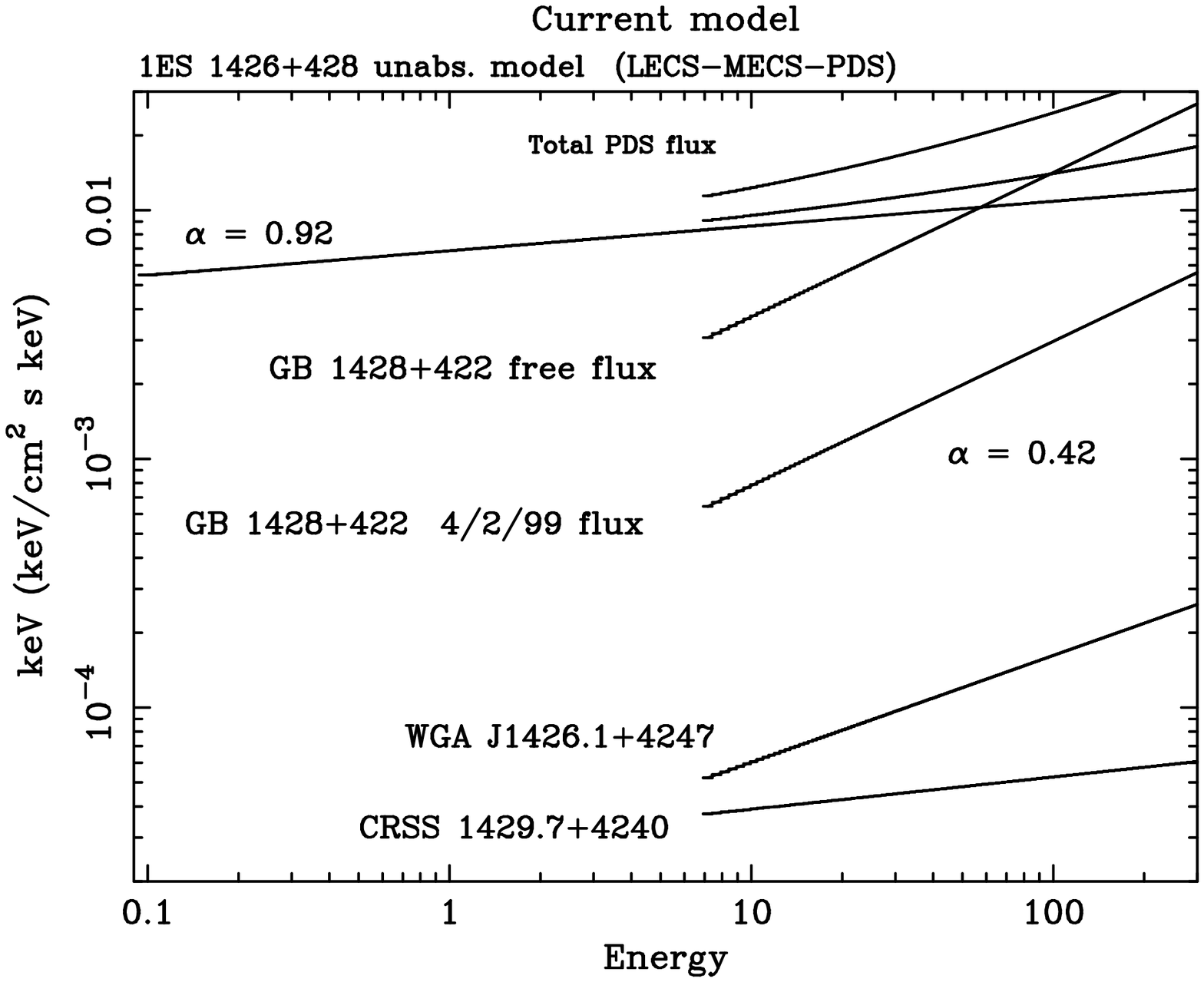, width=8cm, height=10cm }}\\
\end{tabular}
\vspace*{-1.3cm}
\caption[]{\footnotesize Left panel: LECS+MECS single powerlaw fits to 
1ES 1426+428 data, with two different levels for GB 1428+422 flux.  Right panel:
the models used for PDS data.}
\end{figure*}

\section{Results}
The main results for all sources are presented in Table 1.
All have been detected in the PDS band.
For three of them the spectrum is best fitted  with a convex broken power-law:
this locates the peak of the synchrotron emission in the X--ray band,
around 1-4 keV, thus confirming the ``extreme" nature of these sources.
The spectrum of 1ES 1426+428 is  instead well fitted by a single powerlaw,
with a flat spectral index 
($\alpha=0.92$) up to 100 keV.
{\bf This constrains the synchrotron peak to lie near or above 100 keV}.
Such high values of the synchrotron peak frequencies, flagging the presence of
high relativistic electrons, make these sources 
good candidates for TeV emission through the Inverse Compton mechanism.

\begin{table*}[t]
\footnotesize
\caption{LECS+MECS best fits parameters}
\begin{tabular}{llllllll}
\vspace*{-1mm}\\
\hline
\hline
\vspace*{-2mm} \\
Source,  & N$_{\rm H}$ & $\alpha_1$ & E$_{break}$ & $\alpha_2$ & $F_{1keV}$ &
$F_{2-10}$  & $\chi^2_r$/d.o.f. \\
\vspace*{2mm}
date, model & $10^{20}$ cm$^{-2}$ & & keV & & $\mu$Jy &  ergs/cm$^{2}$s  &
\vspace*{0mm}\\
\hline
\hline 
\vspace*{-2mm}\\
\multicolumn{8}{l}{\bf 1ES 0120+340 } \\
2/2/99,  BP & $5.2 \,{\it gal.}$  & $0.82^{-0.96}_{+0.26}$ 
 & $1.4^{-0.7}_{+1.2}$ & $ 1.32^{-0.08}_{+0.08}$ & $4.5^{-0.6}_{+2.1}$ &  
$1.3\; \times10^{-11}$ & 0.92/93  
\vspace*{1mm}\\
\hline
\vspace*{-2mm}\\
\multicolumn{8}{l}{\bf PKS 0548-322} \\
%
20/2/99,  BP & $4.2^{-0.9}_{+1.1}$ & $0.91^{-0.16}_{+0.10}$ &
$4.4^{-2.2}_{+1.8}$ & $1.38^{-0.31}_{+0.59}$ & $5.7^{-0.5}_{+0.5}$ &
$2.3\; \times10^{-11}$ & 0.95/82
\vspace*{1mm}\\
\hline
\vspace*{-2mm}\\
\multicolumn{8}{l}{\bf 1ES 1426+428} \\
8/2/99, SP  & $1.5^{-0.3}_{+0.4}$ & $0.92^{-0.04}_{+0.04}$ & --- & --- &
$4.6^{-0.2}_{+0.2}$ & $2.0\; \times10^{-11}$ & 1.00/89
\vspace*{1mm}\\
\hline 
\vspace*{-2mm}\\
\multicolumn{8}{l}{\bf H 2356-309}  \\
21/6/98,  BP  & $1.3 \,{\it gal.} $ & $0.78^{-0.09}_{+0.06}$ & 
$1.8^{-0.6}_{+0.6}$ & $1.10^{-0.05}_{+0.05}$ & $6.2^{-0.5}_{+0.5}$ &
  $2.5\; \times10^{-11}$ & 0.94/35 
\vspace*{1mm}\\
\hline
\hline
\multicolumn{8}{l}{ SP: Single Powerlaw \hspace*{5ex} BP: Broken Powerlaw 
\hspace*{5ex} Errors at 90\% conf. level for 2 par. of interest } \\ 
\end{tabular}  
\end{table*}  

\begin{acknowledgements}
This research has made use of the NASA/IPAC Extragalactic Database (NED) which is operated by the Jet Propulsion
Laboratory, California Institute of Technology, under contract with the National Aeronautics and Space
Administration. We thank the {\it Beppo}SAX Science Data Center for their support in the data
analysis. This research is financially supported by the Italian Space Agency. \\
L.C. thanks the Cariplo Foundation and the Italian MURST for support.
\end{acknowledgements}

\end{document}